\documentclass{ijm}

\begin{document}
\newcommand{\be}{\begin{eqnarray}}
\newcommand{\ee}{\end{eqnarray}}

%
%

\title{STATUS OF CHIRAL DOUBLERS OF HEAVY-LIGHT HADRONS\\
IN LIGHT OF RECENT BABAR, CLEO, BELLE AND  SELEX $D_s$ 
STATES\footnote{Based on 2004 talks at  MESON04 (Krak\'ow, June 4th-June 8th), 
Rencontre des Moriond (La Thuile, March 28th-April 4th),
 PANDA Workshop (Frascati, March 17th-March 19th) and Hanyang Workshop (Seoul, May 6th-May 8th.}
}

\author{\footnotesize MACIEJ A. NOWAK\footnote{nowak@th.if.uj.edu.pl}}

\address{M. Smoluchowski Institute of Physics, Jagiellonian University,
\\
Reymonta 4, 30-059 Krak\'{o}w, Poland
}



\maketitle


\begin{abstract}
We explain the main idea of  the chiral doublers scenario, originating 
from simultaneous constraints of chiral symmetry and of 
heavy quark spin symmetry on effective theories of heavy-light hadrons.  
In particular we discuss chiral doublers for mesons, chiral doublers for 
excited mesons, chiral doublers for baryons and chiral doublers for 
excited baryons.   
We point out the arguments why new states $D_s(2317)$ and  $D_s(2457)$  
 might be viewed as chiral doublers of $D_s$ and  $D_s^*$.
Then we comment on non-strange mesons $D_0(2308)$ and $D_1'(2427)$ 
observed by Belle and Focus, and on $\Theta_c(3099)$ signal observed by H1.   
Finally, we point out that very recent discovery by SELEX
of $D_s(2632)$, if confirmed by other experiments and if spin-parity 
of this state is $1^-$, may be interpreted as a signal for chiral 
doubler of $D_{s1}(2536)$. Such an identification implies    another narrow,
spin-parity $2^-$\, 
 $D_s$ state ca 37 $MeV$ above the new $1^-$, corresponding to 
chiral partner of $D_{s2}$.    

\end{abstract}

\section{New experimental results on open charm}	

During last year, several experiments have reported 
 spectacular discoveries in the domain of charm spectroscopy. In particular:\\
-- BaBar~\cite{BB} has announced new, narrow meson
 $D^{*}_{sJ} (2317)^+$,
decaying into $D_s^+$ and $\pi^0$. This observation was then confirmed by
CLEO~\cite{CLEO}, which also noticed another narrow state,
 $D_{sJ}(2463)^+$, decaying
into $D_s^*$ and $\pi^0$. Both states were confirmed
by Belle~\cite{BELLE}, and finally,
the CLEO observation was also confirmed by BaBar~\cite{BABAR2}.
The 2317 state was also observed by Focus~\cite{FOCUS}.\\
-- Belle provided  first evidence for two new, broad states $D_0^*$ $(2308\pm
17\pm 15\pm 28)$ and $D_1^{'}$ $(2427\pm 26 \pm 20 \pm 17)$
\cite{Bellenonstrange}. Both
of them  are approximately $400$ MeV  above the usual $D_0,D^{*}$
states and  have opposite
to them parity.\\
-- Few days ago, Selex has announced a new, surprisingly narrow  state
   $D^+_{sJ}(2632)$~\cite{SELEXD}, with unusual decay properties.\\ 
-- New results on charmed baryon spectroscopy appeared lately:
   Selex  confirmed   
   doubly charmed baryons~\cite{SELEX}, H1 experiment at DESY has 
   announced~\cite{H1DESY} a
signature for
   narrow charmed {\it pentaquark} $\bar{c}udud$ at mass 3099 MeV.

   The above states  and in particular the decay patterns
of all these  particles challenged   standard estimations
based on quark potential models  and triggered   a renewal 
of interest on charmed hadrons spectroscopy 
among several theorists~\cite{ALLNEW,recsum}.

An appealing possibility is that the presence of several  above states
is the consequence of so-called chiral doublers scenario,
theoretically anticipated~\cite{us92,bh93} already a decade ago.

\section{Chiral doublers scenario}

Quarks with different flavors have very different masses: current masses of 
$l(ight)=u,d,$ and $s$ quark ($1.5-4,4-8,80-130$ MeV) are smaller 
than the fundamental constant of QCD (in $\overline{MS}$ scheme),
 $\Lambda_{QCD}=220 \pm 20MeV$, 
whereas current masses of $h(eavy)=c,b,$ and $t$ quark
($1150-1350,4100-4400,174000\pm 5000$ MeV) 
are considerably heavier than $\Lambda_{QCD}$. 
In order to understand the dynamics of strong interaction, it is tempting to 
consider following limits:\\ 
$\bullet$ $m_l/\Lambda_{QCD} \rightarrow 0$\\
$\bullet$ $\Lambda_{QCD}/m_h \rightarrow 0. $\\
First limit (massless light quark limit) is known as a chiral limit. 
In this case an exact chiral symmetry of the QCD interactions is spontaneously broken. Vacuum state is respecting only vector 
part of the symmetry, whereas axial symmetry is broken, 
resulting in massless Goldstone excitation for each ``direction'' 
of the axial symmetry group. On top of this effect explicit breaking of the 
chiral symmetry (due to the finite masses of light quarks) takes place,
 shifting the massless pions to 140 MeV, massless kaons to 495 MeV etc.  
The breakdown of chiral symmetry leads to a well known wealth of 
predictions on low-energies processes, based on organizing the amplitudes 
in powers of the light meson momenta (low energy theorems, 
chiral Ward identities, chiral perturbation theory etc.)

The second limit (infinite heavy quark mass limit or Isgur-Wise~\cite{IW} 
limit)
 is equally illuminating. 
In this limit, dynamics of the heavy quark becomes independent of its spin 
and mass. As one of the  consequences of such limit, 
the masses of the pseudoscalar $0^-$ and vector
$1^-$ mesonic states including heavy quark become degenerate. 
Again, the systematic expansion in $1/m_h$ is possible, 
establishing the principles of heavy quark effective theory. 

The case of heavy-light mesons, as the simplest heavy-light hadrons,  
is particularly interesting. It is natural, from the aforementioned arguments,
 to 
consider their dynamics as simultaneously 
constrained by {\em two } above limits. After understanding these constraints, 
one can address the issue
of finite light quark masses and/or finite $1/m_h$ corrections.   

The chief observation made in~\cite{us92,bh93} was that a consistent 
implementation of the spontaneous breakdown of the chiral symmetry for light
quarks and Isgur-Wise symmetry for heavy quarks requires 
in addition to known $(0^-,1^-)$ heavy mesons, new heavy-light 
{\em chiral partners}, separated in the Isgur-Wise limit by the finite, small
split originating solely from the spontaneous breakdown of the chiral symmetry. 
The similar doubling was expected 
for heavy-light baryonic states as well for  hadronic excitations~\cite{nz94}. 
This prediction was in contrast to traditional heavy-light spectroscopy, 
which was much based on Coulomb bound states alike heavy-heavy systems,
and was not taking into account the possibility that the 
constraints of chiral symmetry may be so manifest 
at the level of $c$ and $b$ quark physics. 
 
\begin{figure}[h]
\centerline{\epsfxsize=90mm \epsfbox{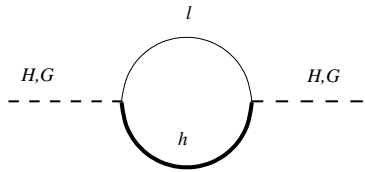}} \vskip -1.5cm
\caption{One-loop contribution  to 2-point $\bar{H}H$, $\bar{G}G$ functions.
Here $l$ stands for light quark and $h$ for heavy quark.}
\label{fig1}
\end{figure}

From the point of view of the chiral symmetry, the chiral doubling 
for heavy-light systems is a fundamental phenomenon, 
representing a pattern of strong 
interaction. Let us present a simple argument, why the presence of doublers 
is not a puzzle.
To one-loop approximation, the order $m_h^0$ contribution to the
mass of the heavy-light meson comes  from the diagram shown in
Fig.~\ref{fig1}. The propagator of the heavy quark $(p\!\!\!/-m_h)^{-1}$
becomes $v\!\!\!//(vk)$, after decomposing the heavy quark momentum into 
$p=m_h v+k$ with four-velocity $v$. The light (massless) quark, 
when propagating through the non-trivial vacuum, dresses itself, and acquires 
a constituent mass $\Sigma$. 
Let us  introduce, after~\cite{GEORGI} a following notation 
for the degenerate in the IW limit 
vector-pseudoscalar, $(0^-,1^-)$ state:
\be 
H=
\frac{1+v\!\!\!\!/}{2}(\gamma^{\mu}D^{*}_{\mu} +i\gamma_5 D)\,\,
\ee
with a transverse vector field, i.e. $v\cdot D^*=0$.
Then the main contribution from the diagram shown 
on Fig.~\ref{fig1} reads
\be 
 m_H {\rm tr} \bar{H} H \sim {\rm Tr} \left( P_l\,\frac{ \Sigma}{Q^2-\Sigma^2}\, H
P_h\frac{\rlap/{v}}{v\cdot Q}\bar{H}\right)\,\,. \label{x2} \ee
where
the trace (tr)   is over flavor and spin, trace (Tr) includes 
additionally integrating over momenta circulating within the loop and $P_l$ and $P_h$ are 
projectors for light and heavy quark propagators (note that $H$
 and its conjugate $\bar{H}$ mixes heavy-light quarks). 

Let us consider now the similar diagram, but for states with opposite parity.
We introduce a natural notation,
\be
G=
\frac{1+v\!\!\!/}{2}(\gamma^{\mu}\gamma_5\tilde{D}^{*}_{\mu} +
i\tilde{D})\,\,.
\ee
for degenerated in the IW limit 
scalar-pseudovector multiplet $G$. We denoted the doublers $(0^+,1^+)$ by 
tilde. 
Then, the mass contribution reads 
\be 
m_G {\rm tr}\, G \bar{G} \sim {\rm Tr} \left( P_l\,\frac{ \Sigma}{Q^2-\Sigma^2}\,
 G
P_h\frac{\rlap/{v}}{v\cdot Q}\bar{G}\right)\,\,. \label{x3} 
\ee
Note that the difference for chiral masses origins 
from the additional $\gamma_5$  in the definition of the
fields $H$ and $G$, (note $\gamma_5^2=1$), 
in other words from the parity assignment.
The range of integration over 4-momentum $Q$
under trace (Tr)   is $0<Q<\Lambda_{UV}$ where $\Lambda_{UV}$ is an
ultraviolet cut-off.
Since
$H\rlap/{v}=-H$ and $G\rlap/{v}=+G$, the result is a split between
the heavy-light mesons of opposite chirality. This unusual
contribution of the chiral quark mass stems from the fact that it
tags to the {\em velocity} $H\rlap/{v}\bar{H}$ of the heavy field
and is therefore sensitive to {\em parity}.
The reparametrization invariance (invariance under velocity
shifts of the heavy quark to order one) introduces mass shifts that are
parity insensitive to leading order in $1/m_h$~\cite{nz94}.
The  $HG$-mass difference is dictated by the spontaneous breaking
of chiral symmetry:\\
 {\bf i)} the light quark contributes a mass
shift of order of an induced cut-off dependent constituent mass
$\Sigma$;\\
 {\bf ii)} interaction is  repulsive in the scalars (no
$i\gamma_5$) and is attractive in the pseudoscalars (with
$i\gamma_5$), so the mass of $G$ goes $up$ and the mass of $H$ goes $down$. \\
In this limit, the spontaneous breakdown of chiral
symmetry enforces the mass relation~\cite{us92}
\be m(\tilde{D}^*)-m(D^*) = m(\tilde{D})-m(D)= m_G - m_H \sim O(\Sigma)
\label{split}
 \ee
If we would restore ( e.g. in Gedankenexperiment) the   chiral symmetry,
i.e. put $\Sigma=0$ 
both chiral copies  will be degenerated. In real world, 
chiral symmetry is broken, as well as IW symmetry is not exact. 
As a consequence, spin 0 and spin 1 states are split by 
the $1/m_h$ effects, and resulting both pairs $(0^-,1^-)$
and $(0^+,1^+)$ are also split by the spontaneous and explicit breakdown 
of the chiral symmetry. This split may be viewed 
(at least in the chiral limit) 
as  an {\it order parameter} for spontaneous breakdown of the chiral symmetry.

Since chiral doublers scenario is based solely 
on patterns of QCD symmetries in the chiral and IW limit, one could 
expect a generic,  model independent manifestation  of  this phenomenon, 
alike the case of low-energy theorems for the light-light systems. 
Indeed, this is the case. Goldberger-Treiman relations for heavy-light
mesons, first written by~\cite{bh93} and extended  for finite 
heavy and finite light masses in ~\cite{usnew} constitute 
model-independent, low energy QCD theorems
for heavy-light mesons. They explicitly demonstrate the appearance of 
small scale of pion decay constant $f_\pi \sim 100 MeV$ together 
with large mass scales of the $D$ or $B$ mesons. This small scale, magnified 
by the ratio of axial coupling constants, yields typical split of order of constituent mass of the light quark, i.e. $350 MeV$.

\subsection{``D-cubes''}

\begin{figure}
\parbox{\textwidth}
{\centerline{
\includegraphics[width=.48\textwidth,height=.37\textwidth]{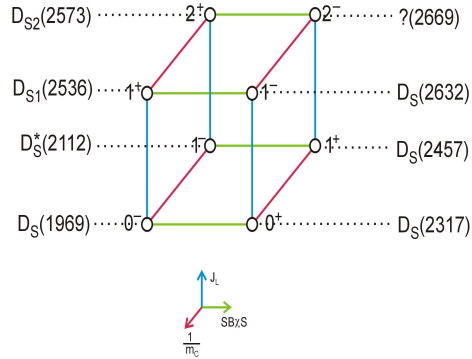}
}}
\caption{Cube  representing {\em schematic} (e.g. the units in the  
upper and lower plaquettes are different) classification of chiral 
doublers. Labels correspond  to the  case of $c\bar{s}$ mesons. Selex
signal $D_s(2632)$ is interpreted as an excited doubler, see text. 
}
\label{figbox}
\end{figure}

One can visualize the consequences of the chiral doubling 
 for mesons in the form of the cartoon, 
see  Fig.~\ref{figbox}. 
 The three-dimensional ``cube'' is aligned along  three ``directions'':\\
- chiral symmetry breaking (horizontal, green)\\
- Isgur-Wise symmetry breaking (skew, red)\\
- total {\em light} angular momentum (vertical, blue).

The corners of the cube represent generic $h\bar{l}$ mesons, i.e. 
we expect similar ``cubic'' patterns  for 
$c\bar{s}$, $c\bar{u}$, $c\bar{d}$, $b\bar{s}$, $b\bar{u}$, $b\bar{d}$ mesons.
   
We  focus on $c\bar{s}$ states, i.e $D_s$-cube. 
Lower left rung represents
known pseudoscalar $0^-$  $D_s(1969)$ and vector $1^-$ $D_s^{*}(2112)$,
belonging to $j_l=1/2$ light angular momentum representation. 
The splitting between them (143 MeV) is an $1/m_c$ effect and is 
expected to vanish 
in infinitely heavy charm quark limit, i.e. both particles would have form 
the $H$ multiplet. The upper left rung corresponds to $j_l=3/2$ representation,
i.e. $1^+$ and $2^+$ {\em excited} multiplet. Here $D_{s1}(2536)$ and 
$D_{sJ}^{*}(2573)$ are the candidates, separated by 
(smaller for excited states, here only 37 MeV)
$1/m_c$ origin mass splitting. 
This ``left plaquette'' 
 of the $D_s$-cube represents  the standard, ``pre-BaBarian''
charmed meson spectroscopy. 

The phenomenon of  chiral doublers implies the appearance  
of the {\em right plaquette}.  
First, we expect two chiral partners for $D_s$ and $D_s^*$, representing right
lower rung.
Here newly discovered $D^{*}_{sJ}(2317)$ and $D_{sJ}(2460)$ are 
the candidates for the $(0^+,1^+)$ scalar-axial $G$ multiplet. 
The averaged splitting for $(0^+,0^-)$ and the averaged splitting for 
$(1^+,1^-)$ are $349.2\pm 0.8$ and $346.8+1.1$, respectively, 
i.e. almost identical, as predicted a decade ago~\cite{us92,bh93}.
It is intriguing that the combined effects  of the finite light mass
 and finite 
charm mass are so small, that both chiral shifts are so close
 to each other already at the level of the mass of the charm quark.
Naturally, the splitting within the $G$ multiplet, i.e. between the masses
of the new BaBar state and CLEO state, is identical  to the 
splitting between the $(1^-,0^-)$ pair.
 
The narrowness of the new states is basically the consequence of the 
kinematic constraint, as pointed by~\cite{beh}: since the chiral split 
is smaller than the mass of the kaon, these states live longer. 
On top of this effect, the isospin conservation most probably 
forces the pionic decay 
via  virtual $\eta$ decay, suppressing the rate even further~\cite{beh}.
Electromagnetic transitions, estimated on the basis of chirally
doubled Lagrangians in~\cite{beh,cf,mehen} are also in agreement with 
the experimental 
data. Chiral Ward identities 
additionally constraint the amplitudes of the pionic decays 
for the $H$ multiplets,  $G$ {\em and} for  $G-H$
 pionic transitions~\cite{usnew}.

On the basis of the chiral doublers scenario, we would also expect 
the chiral partners for the excited $j_l=3/2$ multiplet, i.e. 
{\em new} chiral pair $(1^-,2^-)$~\cite{nz94}. 
Alternatively, this pair could be 
also viewed as the $j_l=3/2$ excitation of the BaBar-Cleo $(0^+,1^+)$ 
multiplet. 
The states within this new multiplet would be separated by similar $1/m_c$
split, like the split between $D_{s2}$ and $D_{s1}$, i.e. by 
37 MeV. A crucial question is how large is  the chiral split 
for the excited states, 
is it also equal to 350 MeV alike the chiral split for the $j_l=1/2$
plaquette or is different? One can try to get some insight 
using the construction for effective chiral action for excited 
mesons~\cite{nz94}. In particular, we 
obtained the formulae for chiral mass shifts for excited mesonic states, 
for $D$ type mesons. After  tuning the ultraviolet cutoff to recover 
the experimental value of 350 MeV  for the chiral split of ground states,
we obtained~\cite{usnew} substantially smaller chiral shift for the 
excited states,
approximately half of the value of the shift for $j_l=1/2$ multiplet
(170 MeV), i.e 
we placed the new pair above 2700 MeV, at $2720$ and $2760$ MeV,
for $1^-$ and $2^-$, respectively.  
The fact that excited states are less sensitive to the effects  of the 
QCD vacuum is not totally unexpected, see e.g.~\cite{GLOZMAN}. 
Of course, the precise value of the chiral shift for  the excited doubler
can be provided only by an experiment. 
It is tempting   to speculate that  the very recent signal reported by 
SELEX~\cite{SELEXD} is a  $1^-$ doubler of $D_{s1}$~\cite{TED}, 
if the state is confirmed and its spin-parity is indeed $1^-$.  
Then the chiral shift for excited strange charmed mesons would be of order of 
100 MeV only. If indeed this is the case, 
a natural expectation in the chiral doubler scenario is 
the presence of the chiral doubler for $D_{s2}$ state as well,  i.e. 
one would expect new, $2^-$ state within few MeV around 2669
MeV, possibly in  $D_s^* \eta$ channel, to  follow the pattern of the decay
of other doublers. 
The presence of such state, completing the identification 
of corners of the ``$D_s$-cube'', could be viewed as a strong argument in favor of the chiral doublers scenario.  


\subsection{D-cube,  $B_s$-cube and $B-cube$}
D-cube construction is generic. 
Left and right plaquettes are chiral copies, 
front and back plaquettes become degenerate in infinite mass of the 
heavy quark and the lower and upper plaquettes are separated by the 
 excitation of total {\em light} angular momentum $j_l$. 
We briefly note  the possibility of even higher angular excitations, 
i.e. additional $j_l=5/2$ plateau, pointing only that first such states
$(2^-,3^-)$ may naturally appear above 3 GeV for $D(D_s)$ mesons.    

Let us consider  non-strange charmed mesons.
Left plaquette is formed by known 
non-strange charmed mesons  i.e. $D(1865)$, $D^{*}(2010)$,
$D_1(2420)$ and $D_2(2460)$. 
Here two states from Belle, $D_0^{*}(2308)$ and $D_1^{'}(2427)$ 
are natural candidates for  lower right rung of the D-cube, i.e.  
for the chiral doublers
of $D(1825)$ and $D^*(2010)$. There are however broad, since neither
kinematic nor isospin restrictions apply here, contrary to their
strange cousins.  
The precise value of  the chiral shift is still an open problem, 
due to the experimental errors and systematic difference between the 
Focus~\cite{FOCUS} and Belle~\cite{Bellenonstrange} signals.  
We would like to mention, that 
the fact that   chiral mass shift 
 seems to be equal of even larger 
 for the 
non-strange mesons than for the strange ones, is not in contradiction
with certain models of spontaneous breakdown of the chiral 
symmetry~\cite{usnew}, although other models make 
opposite prediction~\cite{fajfer}.  We should also mention, 
that the masses (modulo above 
experimental uncertaintities) of these two states can be also 
understood in the quark model.

Let us move towards the $B_s$ and $B$ mesons. In this case,  the chiral
doubling should be even more pronounced for bottom mesons, since
the $1/m_h$ corrections are three times smaller, i.e. the skew-symmetric (red)
edges of the cubes are three times shorter, for $J_l=1/2$ and 
$J_l=3/2$ states, correspondingly. For $m_s=150$
MeV, we expect~\cite{usnew} the chiral partners of $B_s$ and $B_s^*$ to be 323
MeV heavier, while the chiral partners of $B$ and $B^*$ to be 345
MeV heavier, i.e. close to predictions in~\cite{beh}.  
We note that any observation of chiral doubling for $B$
mesons would be a strong validation for chiral doublers  proposal. 
For several recently proposed alternative scenarios for new states
(multiquark states, hadronic molecules, modifications of quark
potential, unitarization) a repeating pattern from charm to bottom
seems to be hard to achieve without  additional assumptions.

\subsection{Chiral doublers for baryons}

In two next  sections, we briefly discuss the extension of the chiral doublers
scenario for all baryons, including the exotic states
(pentaquarks). To avoid any new parameters, we simply view baryons
as solitons of the effective mesonic Lagrangian including {\em
both} chiral copies of heavy-light mesons, 
a point addressed  already in~\cite{us92} and recently reanalyzed 
in~\cite{npsw}.  We are working in large
$N_c$ limit, which justifies the soliton picture, and large heavy
quark mass limit, where we exploit the Isgur-Wise symmetry. This
approach could be viewed as a starting point for including $1/m_h$
corrections from the finite mass of the heavy quark, explicit
breaking of chiral symmetry, etc.

The description of baryons as solitons of the mesonic Lagrangians
has a long history. Original Skyrme~\cite{SKYRME} idea was
elaborated by Witten~\cite{WITTEN}, and Adkins, Nappi and
Witten~\cite{WAN} for $SU(2)_{flavor}$ with enormous success and
hundreds of followers. 
In
~\cite{JMW} it was pointed out how to adapt the soliton scenario
for heavy flavors, i.e. how to use effective chiral Lagrangian 
for light sector and simultaneously 
respect the Isgur-Wise symmetry. The main idea was that 
the SU(2)-flavored   soliton binds  the degenerate in the IW limit pair of a
pseudoscalar and a vector, i.e. $D$ and $D^*$. Charmed hyperons
emerge therefore as bound states of $D$ and $D^*$ in the presence
of the SU(2) Skyrmion (soliton) background. 
Isgur-Wise symmetry at the baryonic level
results in the  degeneration of spin 1/2 and 3/2 multiplets.
Schematically, charmed hyperon of positive parity comes as
\be
 {\rm chiral\,\, soliton} + H\,{\rm multiplet} \rightarrow {\rm charmed\,\, hyperon}
\ee
It is natural to generalize this idea for chiral doublers, i.e 
to obtain charmed hyperons of negative  parity as bound states 
of SU(2) soliton and chiral doublers of $D$ and $D^*$, ie. 
\be
{\rm chiral\,\, soliton}  + G\,{\rm multiplet} \rightarrow 
{\rm chiral\,\, doubler\,\, of\,\, charmed\,\, hyperon}
\ee
Explicitly, we get for both copies (for details and references
see~\cite{npsw})
 \be
M&=&M_{soliton } +m_D -3/2 g_H F'(0) +a/I_1 \nonumber
\\
\tilde{M}  &=& M_{soliton} +m_{\tilde{D}} -3/2 g_G F'(0) +a/I_1
 \ee
 where $M_{soliton}$ is the $O(N_c)$ classical mass of the Skyrmion,
$m_D=(3M_{D^*}+M_D)/4$ is the averaged over heavy spin 
mass of heavy-light mesons,
$m_{\tilde{D}}$ is similar mass for the chiral mesonic doubler with parity
$(0^+,1^+)$, $g_H$ is the axial coupling constant responsible for the $D^*$
decays into a $D$ and a pion, similarly $g_G$ is the corresponding 
axial coupling for the doublers, and the inverse of moment of inertia
of the Skyrmion $1/I_1$ provides the splitting between the various
isospin states ( e.g. for isosinglet $a=3/8$). 
We follow here the conventions of~\cite{OPM}. 
It is of primary importance that, despite the additional
$\gamma_5$ in the definition of the $G$ field
 both mass formulae have the same functional form 
of  $M$ for $H$ and $\tilde{M}$ for $G$. It happens  due to the fact, 
that the term mixing $G$ and $H$ multiplets vanishes for the 
Skyrmion configuration~\cite{npsw}. Hence both
parity partners emerge as $H$ and $G$  bound states in the SU(2)
solitonic background. The mass difference comes: 
first, from meson mass difference $m_{\tilde{D}}-m_D$;  second, 
 from the difference of the coupling constants
$g_G-g_H$.  Using recent Belle data~\cite{BELLE}, {\em i.e.} $0^+$
candidate  $D_0^*$ $(2308\pm 17\pm 15\pm 28)$ and $1^+$ candidate
$D_1^{'}$ $(2427\pm 26 \pm 20 \pm 17)$, we get
$M_{\tilde{D}}=2397$ MeV, unfortunately with still large errors.
Comparing the mass shift between
the lowest $\Lambda_c$ states of opposite parities,
$\Lambda_c(1/2^+, ~2285)$ and $\Lambda_c(1/2^-,~2593)$ we arrive
at chiral baryonic shift $\Delta_B$ = 310 MeV. Similarly, $\Xi_c(1/2^+,~2470)$ and
$\Xi_c(1/2^-,~2790)$ give $\Delta_B$ = 320 MeV.
These numbers suggest, that indeed the leading effect for chiral doubling 
of baryons comes from the chiral mesonic shift, on top of which 
one has to add the smaller effects of different 
axial couplings for two copies. Similar effect is expected for double-heavy
baryons of opposite chiralities. 

\subsection{Chiral doublers for exotic states}

The above formalism allows easily an incorporation of exotic states. 
In the fervor of ongoing discussion on pentaquarks, the issue of heavy 
pentaquarks is far from being academic.
Note, that we may consider the possibility of chiral soliton binding the 
anti-flavored heavy meson ($\bar{c}l$), resulting in bound state
with minimal content of four light quarks and one heavy antiquark. 
Model-dependent  calculations~\cite{OPM} show that the binding in this case 
is three 
times weaker, 
predicting the value of the mass of  isosinglet charmed pentaquark 
with spin-parity $1/2^+$ to be 2700 MeV. 
Repeating this reasoning for the case of chiral soliton binding the 
anti-flavored chiral doubler leads us to the  mass formula for isosinglet
heavy pentaquark of opposite parity
\be
\tilde{M}_5&=& M_{soliton} +m_{\tilde{D}} -1/2 g_G F'(0) +3/(8I_1)
 \ee
Combining above formulae we get the value of chiral shift between 
the pentaquarks expressed in terms of chiral shift for mesons and 
ground state baryons of opposite chiralities~\cite{npsw}, i.e.   
 \be 
\Delta_P=\frac{\Delta_B +2 \Delta_M}{3}\ \sim 350 \pm 60 MeV.
 \ee
where we inferred the 
shift of the opposite parity heavy charmed mesons from very recent
Belle~\cite{Bellenonstrange},  $\Delta_M=425$~MeV
unfortunately with still large errors. This  gives  the
mass of the chiral doubler of the pentaquark as high as $3052 \pm
60$~ MeV.
Recently, H1~\cite{H1DESY} noticed a narrow signal at 3099 MeV interpreted as
charmed pentaquark (although this signal was not  
confirmed by other experiments).  
We dare~\cite{npsw}  to interpret the recent H1
state~\cite{H1DESY} as a parity partner $\tilde{\Theta}_c$ of the
yet undiscovered isosinglet pentaquark $\Theta_c$ of opposite
parity and $M_5\approx 2700$~MeV, i.e. even below the strong decay threshold.
 Chiral doublers  scenario offers also a hint how 
to understand the narrowness of the H1 state. The natural channel for the
decay of this state into a nucleon and {\em parity partners} of
the standard $D (D^*)$ mesons is kinematically blocked.
De-excitation of $\tilde{\Theta}_c$ into $\Theta_c$ and a pion is
isospin forbidden and to $\eta^0$ kinematically blocked.
The only way the decay process may proceed, is a chiral
fluctuation of a bound $\tilde{D}$ into $D$ by virtual interaction
with a pion from the nucleon cloud. That requires, however,
spatial rearrangement, since the $D$ meson must be in a partial
wave of opposite parity with respect to the partial wave of
$\tilde{D}$ . Hence the overlap of the $\tilde{D}$-soliton bound
state wave function with the one of the  $D$-soliton is expected
to be small.
 Above argument shows, 
that  the surprisingly heavy mass of
the charmed pentaquark combined with its narrow width, 
if confirmed by other experiments,  
may be qualitatively explained in
the chiral doubler scenario.
Obviously, if heavy pentaquarks do not exist, one does not expect 
the existence of their chiral partners.

\section{Summary}
Chiral symmetry plays crucial role in understanding the properties of 
light hadrons, and it is exciting that the consequences of spontaneous 
breakdown of chiral symmetry may be so dramatic even at the level 
of  charmed and bottomed hadrons. We hope, that the renaissance
of the charm spectroscopy, causing   present excitement 
in both the experimental and theoretical hadronic physics community,
and several on-going and planned new experiments, 
will trigger a major effort to explore what the new states discussed 
in this talk teach us about  non-perturbative QCD.

\section*{Acknowledgments}

This talk is based on  work done in collaboration with 
Mannque Rho, Ismail Zahed, Micha\l{} Prasza\l{}owicz, Mariusz Sadzikowski
and Joanna Wasiluk~\cite{us92,nz94,npsw,BOOK}. 
I am  grateful to the organizers for  the invitations,  
in particular to Paola Giannoti (PANDA), to Jean Tran Thanh  Van and 
 Etienne Aug\'{e}
(Moriond), to Su Houng Lee and  Hyun Kyu Lee (Hanyang),  
and to Reinhard Kulessa, Andrzej Magiera and  Antoni Szczurek (MESON04).


\end{document}